# On Democracy in Peer-to-Peer systems

Ranieri Baraglia, Patrizio Dazzi, Matteo Mordacchini
Information Science and Technologies Institute "A. Faedo"
Italian National Research Council (CNR)
Pisa, Italy
{r.baraglia, p.dazzi, m.mordacchini}@isti.cnr.it

Laura Ricci, Luca Alessi
Department of Computer Science
University of Pisa
Pisa, Italy
ricci@di.unipi.it, alessi.it@gmail.com

*Abstract*—The information flow inside a P2P network is highly dependent on the network structure. In order to ease the diffusion of relevant data toward interested peers, many P2P protocols gather similar nodes by putting them in direct contact. With this approach the similarity between nodes is computed in a point-to-point fashion: each peer individually identifies the nodes that share similar interests with it. This leads to the creation of a sort of "private" communities, limited to each peer neighbors list. This "private" knowledge do not allow to identify the features needed to discover and characterize the correlations that collect similar peers in broader groups. In order to let these correlations to emerge, the collective knowledge of peers must be exploited. One common problem to overcome in order to avoid the "private" vision of the network, is related to how distributively determine the representation of a community and how nodes may decide to belong to it. We propose to use a gossip-like approach in order to let peers elect and identify leaders of interest communities. Once leaders are elected, their profiles are used as community representatives. Peers decide to adhere to a community or another by choosing the most similar representative they know about.

## I. INTRODUCTION

What do people want to know? This is a very hard question to answer, probably it is impossible to give an universal answer due to different interests different groups of people are interested to. It is possible that who is reading this paper wants to know about the last Steve Jobs keynote, whereas other persons want to gather all the available information about the last car released by BMW. These different willings lead people to build specific connections (or better, strike up friendships) with other people on the basis of common shared interests. These "connections" make possible the human social behavior of spreading knowledge by exchanging information between people that are in direct contact, in other words the human activity also known as *word-of-mouth*. This well known phenomena has inspired the *automating the word-of-mouth* [12] approach. A very effective solution that aims to exploit collaborative and social information for filtering data. In particular, in its first formulation has been exploited to filter information such as music albums and artists. However, after the publication of [12], several other systems (mainly centralized) like search engines (e.g. Yahoo!, Google or Bing), as well as some online stores (e.g. Amazon, eBay) have been exploited the concept behind the [12] paper. Indeed, in a world where there is a growing need to rapidly access and be aware of many types of distributed resources like Internet pages, shared files, online products, news and information, a flexible and efficient mechanism as [12] is, exhibits relevant social and economic impacts. However, despite its effectiveness, scalability concerns are always related with these type of approaches. Along the years, in order to address these scalability issues, several approaches has been proposed. Beyond the obvious parallelization of the approach, that in this case remains centralized. One of the most promising alternatives is the P2P approach. Peer-to-peer solutions scale well to large numbers of peers and deal gracefully with system dynamism, whereas centralized systems need expensive and complex techniques to ensure continuous operation under node and link failures. P2P systems are implemented through the collaboration of the peers without needing any centralized authority that would store all the user data.

Among the different approaches based on the peer-to-peer paradigm, the ones exploiting the Gossip ([5], [6]) technique looks very promising. These protocols stems from their ability to reliably pass information among a large set of interconnected nodes, even if the nodes join and leave the system. Furthermore, as the name suggests, these protocols model in a very close way the behavior of people. Indeed, according to the gossip-based protocol, each node in the network periodically exchanges information with a subset of other peers. The choice of this subset (the peer community) is crucial to the proper dissemination of the gossip. An example exploiting such paradigm is the one we presented in [8]. It proposed an architecture for a system that exploits the collaborative exchange of information between peers in order to allow them to build a proper neighborhood sharing common interests. The ultimate aim of the paper was to allow to each peer to build a neighborhood with which to exchange information, suggestions and recommendations based on the *word-of-mouth* concept. With this approach the similarity between nodes is computed in a point-to-point fashion: each peer individually identifies the nodes that share similar interests with it and stores in its neighborhood the set of closer peers it found in the network so far. This leads to the creation of a sort of "private" communities, limited to each peer neighbors list. This "private" knowledge do not allow to identify the features needed to discover and characterize the correlations that collect similar peers in broader groups. In order to let these correlations to emerge, the collective knowledge of peers must be exploited. In order to avoid the "private" vision of the network, the main problem to overcome is related to how to determine, in a distributed fashion, the representation of a community as well as to decide how nodes may choose to join that community. To address this limitation, in this paper we propose AP2PLE (Asynchronous Peer-to-Peer Leader Election), an approach that allows peers to create explicit communities. The approach is based on the identification and the election of communities leaders. Once a leader has been elected, its profile is used as community representation. It is worth to point out that the role of the leader is just to give to the leaded community a recognizable representation (i.e. the leader is NOT a SuperPeer). By using this representation the non-leader peers can decide to adhere to a community or another by comparing their profile against the leader ones. At the end, they will belong to the community represented by the leader with which they share the highest profile similarity.

The remaining of this paper is organized as follows:

Section II describes the solutions already present in the literature. Section III describes the problem we address in this paper. Section IV presents a detailed description of our solution. Section V presents a step-by-step example describing the algorithm behavior. Finally, Section VI presents conclusions and considerations about this work.

## II. RELATED WORK

In the past several solutions have been proposed for distributing data over an overlay P2P network to enable better resource tracings and better connectivity between similar elements. Such solutions exploit different techniques such as Gossip [4], [15], [8], [5], [6], [2], Semantic Overlay Network [1], Election-based mechanism [7], [10].

In [4] the GosSkip systems is proposed. It is a self organizing and fully distributed overlay that provides a support to data storage and retrieval in peer-to-peer environments. It links objects rather than computational entities in a structure formed by a set of balanced trees. GosSkip is built using a gossip-based protocol that organizes peers so that they form an ordered double-linked list (or ring). In GosSkip a peer is associated with a single item of data, and it is managed by the physical node that published the item. Each peer has a name that describes the semantics of the object to which is associated. The only property needed is that these names follow a total and deterministic order. So, the position of an element is fully determined by its name. For the information distribution, it exploits an epidemic-based protocol that maintains $O(log(N))$ peer states, and with a routing cost of $O(log(N))$.

To associate links to published object can lead to a very large number of connections. This is especially true in networks where the number of objects shared by each user is large. For example, networks for media sharing. Furthermore, using GosSkip it is difficult to search not knowing the exact name which identifies the item you are looking for.

The aim of the study conducted in [1] is to reduce the search time of queries executed in peer-to-peer networks that create the overlay network randomly, and at the same time to maintain a high degree of autonomy of the nodes. The authors propose node connections influenced by content, so that, for example, nodes having many "Jazz" files will connect to other similar nodes. Thus, semantically related nodes form a Semantic Overlay Network (SON). The overlay network is organized in SONS, i.e. a tree where each node consists of clusters of peers. Queries are routed to the appropriate SONs, increasing the chances that matching files will be found quickly, and reducing the search load on nodes that have unrelated content. The SON-based overlay networks have a rigid predefined structure. The overlay network does not allow a dynamic reorganization of itself and leads to have clusters of not semantically related node also when solutions structured in multiple levels that can mitigate this disadvantage are exploited.

The paper [9] proposes the Peer Data Management System (PDMS) as a solution to the problem of large-scale sharing of semantically rich data. A PDMS consists of semantic peers connected through semantic mappings. Querying a PDMS may lead to poor results because of semantic degradation caused by the approximations provided during the construction of semantic mappings, which raises the problem of how to promote a mapping of the network in PDMS. The authors propose a strategy for incrementally maintenance a flexible network organization that clusters together peers, which are semantically related in SONs also guaranteeing a high degree of node autonomy. The lack of a common dictionary between peers may lead to similar content even if not the same, but described with the same concepts. To resolve this heterogeneity nodes with semantically similar concepts are grouped in the same SON. A critical aspect of this solution is the evolution of the interests of a peer that leads in changing the concepts it represents. This aspect introduces overheads in the maintenance of the SONs because changes involve changes in the structure of the network that are made through complex procedures. In fact, changes in a peer concepts triggers a distributed mechanism to reorganize the overlay network that involves to all the neighbor peers belonging to the related SONs.

In [15] a proactive method to build a semantic overlay is proposed. Such approach is structured according to two layers. The top layer exploits the Vicinity [14] gossip-based protocol to optimize semantic lists for searching only. The bottom layer offers a fully decentralized service for delivering information on new events. At this level the Cyclon [13] protocol is used to uniformly and randomly select peer throughout the network. Such peers are then passed to Vicinity to make lists of semantic peers. To form such lists the peer clustering is done in a completely implicit way, i.e. without requiring the user to specify any preferences or to characterize the content of files being shared. The conducted evaluation carried out that such lists are highly effective when searching for content. The construction of these lists through epidemics is efficient and robust, even in the presence of changes in the network. A disadvantage of this solution consists in the fact that the clustering of the concepts could lead to reduce the reachability between peers. Each peer is semantically related only to the similar ones, this can lead to divide peer into compartments isolated from each other. The lack of an overall or distributed structure describing the state of the network does not allow a global routing query.

The paper [14] presents an efficient randomized algorithm for leader election in large-scale distributed systems. The proposed algorithm is optimal in message complexity ($O(n)$ for a set of $n$ processes), has round complexity logarithmic in the number of processes in the system, and provides high probabilistic guarantees on the election of a unique leader. The algorithm relies on a balls and bins abstraction and works in two phases. The main novelty of the work is in the first phase where the number of contending processes is reduced in a controlled manner. Probabilistic quorums are used to determine a winner in the second phase. The protocol is completely probabilistic. Do not choose the best represented, but one in a random order.

In [11] a peer connectivity-based distributed solution to cluster node is proposed. It assumes each node knowledges only its direct neighbors. The protocol allows the entire network clustering around a set of starting nodes. The connection structure of a P2P network is represented by a graph where nodes form the vertices and connections between nodes are the branches of the graph. The algorithm can be extended and used in a weighted graph where the weight associated with a connection between two nodes represents the similarity between the two neighbors. Each node completely ignores the structure of the network. Also consider working on a network of highly dynamic with frequent entry and exit of nodes. A critical aspect of this solution is that the division of the nodes in clusters depends greatly on the choice of the initial nodes from which the algorithm starts. This may determine that the chosen nodes is significantly different from the ideal one (medoid) to represent the cluster. Also, the number of clusters is fixed, they can evolve only in size (insertion/delete of node in/by the network), new cluster are not created to reflect changing in the structure of the network. Moreover, the resulting clusters are all at the same level excluding the ability to model sub-categories as described in SONs.

## III. PROBLEM DESCRIPTION

Networks of entities (e.g. humans, peers, ...) have enlightened to have emerging structures, embodied in the relationships among their participants. The properties of these structures, or communities, are of the most importance to guarantee a fast and efficient communication among entities.

In order to discover those relationships, the main problems are the identification and representation of these structures. Consider a scenario of a set of entities. Every entity has a group of different profiles, each representing a different interest. Generally, one common solution, for every distinct interest, is to partition to whole set of entities into groups showing a good grade of homogeneity.

More formally, given a set $\mathcal{N}_I$ of entities, for the interest $I$, we want to have a partition:

$$\mathcal{N}_I = \{N_1, \ldots, N_s\}$$

where each $N_i$ contains a homogeneous subset of the entities in $\mathcal{N}_I$.

This goal implies to have a tradeoff on the cardinality of such groups. Larger groups could have too little internal correlation between members, whereas smaller groups could be too elitist, failing to catch existing relationships with members of other groups.

Moreover, for each $N_i \in \mathcal{N}_I$, we wish to find a proper label $e_i$ that gives an easy and immediate way to detect the membership to the group and to recognize its content.

All the above issues are make even harder in distributed, dynamic environments like Peer-to-Peer (P2P) networks. This kind of networks is characterized by the lack of global knowledge, a high number of participating peers and a high grade of changes due to join/leave operations performed unpredictably by each node.

In order to face those problems, many approaches try to gather similar nodes by putting them in direct contact. This behavior allows each node to store the most similar other peers it encounters during message exchanges with other nodes. On of the most significant limitations of this approach is that similarity of interests between nodes is computed in a point-point fashion. In other words, each peer individually computes which are the other nodes that may share similar content/interest with it. This leads to the creation of a sort of "private" communities, limited to each peer neighbor list. This "private" knowledge do not allow to give a good identification of the features needed to characterize such communities, since it does not give any recognizable identifier for communities of similar peers, thus failing to reach the goal stated at the beginning of this section.

Thus, a more extended collaboration between peers has to be exploited in order to form a collective agreement over communities and community representatives.

## IV. THE PROPOSED SOLUTION

Our main aim is to let peers spontaneously gather into communities of similar nodes. In order to have a "public" recognition of a community, we wish to create a community label, that is used by each peer to characterizes itself as member of the community and it is then further exploited to communicate, exchange and spread information along the network. We propose to use a gossip-like approach that allows peers to collaborate in order to create the communities and find good identifiers for them. We choose to use the profile of a peer inside each community as the community identifier. These profiles are the ones that are chosen by the other peers of each community as their best representative.

Each peer is associated with a network user. Each peer is characterized by the profile of this user. As described in [3], a profile consists of a number of terms k, where k varies from user to user. The similarity between two peers is the ratio between the number of terms that the two peers have in common and the union of terms of peer profiles.

This approach, due to its simplicity, can be applied to several different types of profiles. Just to name a few, it can be applied to profiles that describe the information associated with web visited pages, tags of multimedia data files, or any profile extracted from text documents by assuming to have a detector of keywords, or alternatively, to profiles made of a list of words specified by the user.

Our proposal exploits a distributed election mechanism that allows both the gathering into communities and the determination of which are the those significant profiles for each community. The gossip mechanism, starting from the "private" vision of each node, after successive interactions and exchange of information with other peers, allows to exploit each peer local knowledge to contribute to the identification of representatives.

The approach presented in this article assumes that each peer belonging to the network has the ability to find his more similar neighbors throughout the system, as described in [15]. To this end, the Cyclon [13] and Vicinity [14] protocols are used to both maintain a connected network and to let similar peers to move toward nearby areas of the network. As a consequence of the above approaches, since each peer retains only a limited number of similar nodes, the network can be described as a directed graph.

In order to build up communities, we propose to use a voting procedure, which leads to select a set of representative peers (leaders). Each peer votes by sending a message. Every expressed vote has associated a TTL (Time-To-Live), at the end of which it is discarded, and it is not counted any more. Each elected representative peer, together with the peers that have contribute to its election and autonomously decide to join it, constitutes a community. A community is characterized by the profile of its leader, that is adopted by each community member as its community identifier.

The proposed voting procedure is structured according to three stages: 1) identification of leader candidates, 2) election of potential leaders, and 3) choices of leaders. The algorithms used are shown from Alg.1 to Alg.5.

---

**Algorithm 3** CandidateSelectionAndVote

1: **procedure** CANDIDATESELECTIONANDVOTE
2:     $Order$ NEIGHBORS $by$ similarity;
3:     $Let$ BESTN = Top-k(NEIGHBORS);
4:     **for all** $b \in$ BESTN **do**
5:        **if** $d(p, b) \leq (1 - neighbor\_threshold)$ **then**
6:           $Send$ Vote $to$ $b$;
7:        **else**
8:           $Break$;
9:        **end if**
10:     **end for**
11: **end procedure**

---

*1) Identification of leader candidates:* This phase aims to identify leader candidates. It consists in a preliminary voting procedure in which each peer votes for carrying out its most similar peers, i.e. its best neighbors in term of profile.

This phase is mainly driven by tree parameters:

- $n\_votes$: the maximum number of neighbors that a peer can vote for;
- $neighbor\_threshold$: the peer similarity value under which two peers are not considered similar, and therefore not selected to be voted.
- $leader\_threshold$: The minimum number of votes to be reached by a peer to be considered as a leader candidate.

Each voting peer arranges its neighbors in decreasing order with respect to the similarity value, and gives a vote to, at most, the $n\_votes$ ones with the highest similarity value, without considering the peers whose similarity is lower than $neighbor\_threshold$.

At the end of this phase, a peer $p$ has received a number of votes given by:

$$\sum_{n \in Nbg_{In}} v_n(p)$$

where:

$$v_n(p) = \begin{cases} 1 & \text{if } p \in \text{top } (Nbg) \text{ of } n \\ 0 & \text{otherwise} \end{cases}$$

Here, $Nbg_{In}$ is the set of nodes for which $p$ has an incoming link connection, $Nbg$ is the usual set of neighbors (i.e. the neighbors with which there exist an outgoing link) and $top(Nbg)$ is a function that selects the first $n\_votes$ most similar peers in $Nbg$. Clearly, for a node $p$ the intersection between $Nbg_{In}$ and $Nbg$ is not void **iff** there exist at least another node $p'$ s.t. $p \in Nbg(p')$ and $p' \in Nbg(p)$.

All the above operations are describe in Alg.3.

*2) Identification of the potential leaders:* This phase is devoted to the identification of potential leader candidates, each peer can give up to $n\_leader$ votes.

---

**Algorithm 4** PotentialLeaderIdentification

1: **procedure** POTENTIALLEADERIDENTIFICATION
2:     $Let$ CANDIDATES $= \emptyset$;
3:     **for all** $n \in \{$NEIGHBORS $\cup \{p\}\}$ **do**
4:        **if** VotesRcvd($n$) $\geq$ LEADERTHR **then**
5:           $Let$ CANDIDATES = CANDIDATES $\cup \{n\}$
6:        **end if**
7:     **end for**
8:     **if** CANDIDATES $= \emptyset$ **then**
9:        $Let$ CANDIDATES = NGHLEADERS
10:        **if** CANDIDATES $= \emptyset$ **then**
11:           $Let$ CANDIDATES $= p$ $ad$ $interim$
12:        **end if**
13:     **end if**
14:     $Let$ L $= \min_{c \in \text{CANDIDATES}} d(p, c)$
15:     $Send$ LeaderVote $to$ L
16: **end procedure**

**Algorithm 1** ActiveThread

1: **procedure** ACTIVETHREAD
2:    **while** $true$ **do**
3:
4:       **if** Timer $t$ expired **then**
5:          $Reset\ t$;
6:          CandidateSelectionAndVote();
7:       **end if**
8:       **if** Timer $t'$ expired **then**
9:          $Reset\ t'$;
10:         PotentialLeaderIdentification();
11:      **end if**
12:      **if** Timer $t''$ expired **then**
13:         $Reset\ t''$;
14:         ActualLeaderElection();
15:      **end if**
16:
17:    **end while**
18: **end procedure**

**Algorithm 2** PassiveThread

1: **procedure** PASSIVETHREAD
2:    **while** $true$ **do**
3:       $Receive$ MSG $as$ MESSAGE
4:       MTYPE = MSG.type;
5:       MTIMESTAMP = MSG.timestamp;
6:       MCONTENT = MSG.content;
7:       MSENDER = MSG.sender;
8:       **if** MTYPE = Vote **then**
9:          $Add$ MCONTENT $to$ VOTESQUEUE
10:                     $with$ MTIMESTAMP
11:          $Send$ VOTESQUEUE.size $to$ MSENDER
12:      **else if** MTYPE = LeaderVote **then**
13:         $Add$ MCONTENT $to$ LDRVOTESQUEUE
14:                    $with$ MTIMESTAMP
15:         $Send$ LDRVOTESQUEUE.size $to$ MSENDER
16:      **end if**
17:    **end while**
18: **end procedure**

The leader votes are assigned to the most similar leader candidate. The potential leader candidates are the ones that received a number of votes higher than the $leader\_threshold$. Thus, usually a peer $p$, identifies the potential leaders by selecting a set of candidates from its neighbors, defined as CANDIDATES$= \{n \in Nbg(p)|VotesRecvd(n) \geq$ LEADERTHR$\}$. Then, it votes a neighbor $L$ as a potential leader using the following function:

$$Vl_p(L) = \begin{cases} 1 & \text{if } L = \min_{c \in \text{CANDIDATES}} d(p,c) \\ 0 & \text{otherwise} \end{cases}$$

Anyway, in order to face same special cases that might occur, the above function needs to be modified according to what is specified in Alg.4.

As a consequence, by acting on the $leader\_threshold$ parameter, it is possible to influence the total number of leaders, and, assuming to keep fixed the other parameters, the cardinality of each community of peers.

*3) The leaders election:* In this phase each peer $p$ elects its actual leaders. At this step $p$ considers also the leader votes it has received from other nodes. At a first sight we can say that a leader candidate $LC$ is chosen by $p$ in the following way:

$$LC = \begin{cases} p & \text{if } p \text{ has the hgst \# of LeaderVotesRcvd} \\ \min_{\substack{l \neq p \\ l \in \text{LCANDIDATES}}} d(p,l) & \text{otherwise} \end{cases}$$

where LCANDIDATES is the set of potential candidates filtered by the votes expressed in the previous phase.

As can be seen, in case $p$ has the highest number of leader votes with respect to its neighborhood, it considers itself as its own leader.

**Algorithm 5** ActualLeaderElection

1: **procedure** ACTUALLEADERELECTION
2:    $Let$ BOSS = NEIGHBORS $\cup$ CANDIDATES $\cup \{p\}$;
3:    $Let$ LCANDIDATES = $\emptyset$;
4:    **for all** $n \in$ BOSS **do**
5:       **if** LeaderVotesRcvd($n$) $\geq$ LEADERTHR **then**
6:          $Let$ LCANDIDATES = LCANDIDATES $\cup \{n\}$
7:       **end if**
8:    **end for**
9:    $Order$ LCANDIDATES $by$ LeaderVotesRcvd;
10:   **if** $p$ = head(LCANDIDATES) **then**
11:      $Set\ p\ as$ ACTUALLEADER;
12:   **else**
13:      $Let$ LC $= \min_{\substack{l \neq p \\ l \in \text{LCANDIDATES}}} d(p,l)$
14:     **if** LC $\notin$ NEIGHBORS **then**
15:        $Set$ LC $as$ ACTUALLEADER;
16:     **else if** $d($LC, LC.$leader) > \epsilon$ **then**
17:        $Set$ LC $as$ ACTUALLEADER;
18:     **else**
19:        $Set$ LC.leader $as$ ACTUALLEADER;
20:     **end if**
21:   **end if**
22: **end procedure**

Otherwise, usually, a potential leader $L$ become actual leader iff, among the peer's neighbor, $L$ is the one that has received the highest number of leader votes. With respect to the previous definition of $LC$, however, if there are two potential leaders received the same number of votes, $p$ chooses its most similar one.

Anyway, there are same other special cases that have to be addressed. When $p$ communicates with its actual leader $L$, and discovers that it has as its own actual leader another node $L'$ that is within a distance $\epsilon$ from $L$, then $p$ changes its decision and adopt $L'$ has its actual leader. In the case when $L$ has $L'$ has its leader, but they are sufficiently separated ($d(L, L') > \epsilon$), $L$ remains as the leader of $p$. The link between $L$ and $L'$ represents a connection among their respective communities, augmenting the spreading of information inside the network.

Another special case occurs when none of the potential leaders of $p$ can be considered its actual leader. Then, $p$ ask to its neighbors for their actual leaders. From the answers received, it selects the one sharing the highest similarity value.

All the above cases and behaviors are described in Alg.5.

At the end, this process leads the peers in the network to spontaneously gather into communities, each of them including all the peer that have chosen the same actual leader.

As before mentioned, each vote has a limited life, exactly as happens in a democracy where a mandate expires after a certain amount of time. Thus, periodically, at predefined interval time, each peer contributes to the election of community leaders. The continuous refresh of information is ensured by the underlying gossip protocols, which are used to put in contact similar nodes and to support an epidemic diffusion of the information, leader votes included.

### A. Split and merge of Communities

As happens to leaders, also communities are subjected to changes. Indeed, communities are not static entities that once formed remain unchanged but are subjected to changes. Beyond the single joins and leaves due to peers that either churn or connect to the network, the two major changes the communities are subjected to potential split and merge.

Anyway, all the operations described in the previous sections are performed by each peer individually, without any form of synchronization with other nodes. The only interaction is the exchange of information, in the form of both exchange of votes and gossip updates (that includes peers' profiles, received votes and leader votes, actual community profiles, etc.). When a sufficient time has passed, each peer independently starts a new flow of votes. At the end of this process it is able to cope with new situations, like the arrival or departure of other peers. The underlying mechanism of Cyclon and Vicinity let $p$ to have an updated situation of similar peers in its neighborhood. Hence, when it will vote again, those new peers will be considered and old or disappeared ones will not be taken into account. At the end, if there sufficient changes in $p$'s neighborhood, it will choose a new community, possibly joining peers of older communities or splitting from its old one to join a new group. Thus, no explicit mechanism to handle joins or splits of communities is required.

## V. THE PROTOCOL IN ACTION

In order to present in a more straightforward way the protocol behavior, in this section a step-by-step description of the protocol execution phases is provided.

Figure 1(a) shows the peer-to-peer overlay network utilized in our example. In the figure, circles represent the peers, arrows represent the relations among peers and the numbers close to the circles indicate the votes that the related peer has received from its neighborhoods in each phase of the protocol. Moreover, in Figure 1(a), (b), (c), (d) and (e) arrows indicate the peers issued and received a vote. As shown in that figure, we assumed to run our protocol on a peer-to-peer overlay network where each peer is connected with three other peers, which are chosen to be the ones most similar to it among the ones belonging to the whole network. The peers similarity is measured by using a function that compares sets of RGB data, each one random generated and representing the profiles of a peer. Such function returns a value indicating how much similar two peers are.

Table I shows for each peer its identification (col. Peer ID), its neighborhood (col. Peer Neighbors), the votes received in the first phase (col. Leader Candidate Votes), the votes received in the second phase (col. Potential Leader Votes) and the identification of its potential leader (col. Potential Leader ID), the votes received in the first and second steps of the third phase (columns Actual Leader votes) and the identification of the its actual leader (col. Actual Leader ID).

In this example the peers similarity was computed as the sum of the absolute values of the differences between the three RGB elements present in each peer profile. Moreover, the protocol's parameters $neighbor\_threshold$, $n\_votes$, $leader\_threshold$, $n\_leader$ were fixed equal to 160, 2, 3 and 1, respectively. We consider two peers to be similar if their distance is less the 160.

Figure 1(b) shows the results of the first protocol phase, in which the leader candidates are chosen. According to the $neighbor\_threshold$ and $n\_votes$ parameter values, each peer votes for its 2 best neighbors, which distance is less than 160. It is worth to note that the node number 9 expresses only one vote, indeed, as can be seen in Table I, only one of its neighbors is distant less than the value indicated by the $neighbor\_threshold$ parameter. The nodes 0, 4, 8, 10, 13, and 19 have received a number of votes greater than the $leader\_threshold$ value, and therefore they are selected as leader candidates.

The second phase is devoted to select the potential leaders. In this, each peer votes at most a number of neighbors candidate leaders equal to the $n\_leader$ value (i.e. 1), and with which it shares the higher similarity. Figure 1(c) shows the results of the second protocol phase. As shown in table I the peers 4, 8, 10, 13, and 19 are selected as potential peers, since they have received at least one vote from its neighbors. The peers 9, and 14 do not have any potential leaders because none of their neighbors is a candidate leader. Moreover, they cannot choose themselves as potential leaders because they received a number of votes less than the $leader\_threshold$ value. These nodes will select its actual leader in the next protocol phase. Moreover, the peers 3, 4, 11, and 13 have more than a neighbor leader candidate. Therefore, they choose as potential leader the leader candidate with which they share the higher similarity. So, the peers 3, 4, 11, and 13 choose as potential peers the peers 8, 13, 4, and 4, respectively. The peers 0, 4, 10, and 19 that have received a number of votes equal or greater than the $leader\_threshold$ value (i.e. 3) by increases their number of votes of a unit elect themselves as potential leaders. This is shown in Figure 1(c).

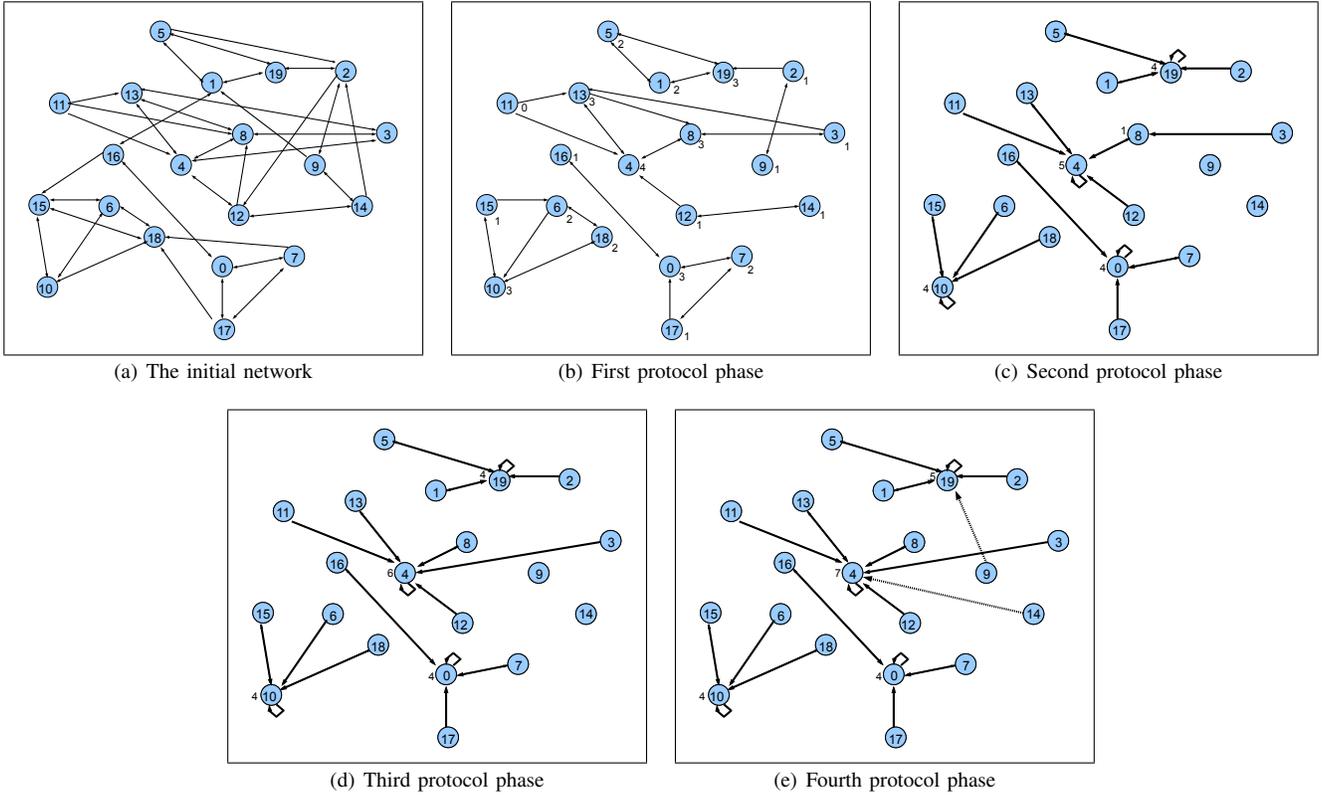

Fig. 1. The protocol phases

TABLE I
PEER VALUES

| Peer ID | Neighborhood | | | First Phase | Second Phase | | Third Phase: 1 step | | Third Phase: 2 step | |
| | Peer Neighbors | | | Candidate Leader Votes | Potential Leader ID | Potential Leader Votes | Actual Leader ID | Actual Leader Votes | Actual Leader ID | Actual Leader Votes |
| --- | --- | --- | --- | --- | --- | --- | --- | --- | --- | --- |
| 4 | id:13 dist: 37 | id: 8 dist: 54 | id:12 dist: 112 | 4 | 4 | 5 | 4 | 6 | 4 | 7 |
| 19 | id: 5 dist: 69 | id: 1 dist: 98 | id: 2 dist: 119 | 3 | 19 | 4 | 19 | 4 | 19 | 5 |
| 0 | id:16 dist: 115 | id: 7 dist: 122 | id:17 dist: 134 | 3 | 0 | 4 | 0 | 4 | 0 | 4 |
| 10 | id:15 dist: 100 | id:18 dist: 109 | id: 6 dist: 112 | 3 | 10 | 4 | 10 | 4 | 10 | 4 |
| 8 | id: 4 dist: 54 | id: 3 dist: 72 | id:13 dist: 87 | 3 | 4 | 1 | 4 | 0 | 4 | 0 |
| 13 | id: 4 dist: 37 | id: 8 dist: 87 | id: 3 dist: 125 | 3 | 4 | 0 | 4 | 0 | 4 | 0 |
| 1 | id: 5 dist: 65 | id:19 dist: 98 | id:16 dist: 167 | 2 | 19 | 0 | 19 | 0 | 19 | 0 |
| 5 | id: 1 dist: 65 | id:19 dist: 69 | id: 2 dist: 188 | 2 | 19 | 0 | 19 | 0 | 19 | 0 |
| 6 | id:18 dist: 25 | id:10 dist: 112 | id:15 dist: 154 | 2 | 10 | 0 | 10 | 0 | 10 | 0 |
| 7 | id:17 dist: 48 | id: 0 dist: 122 | id:18 dist: 219 | 2 | 0 | 0 | 0 | 0 | 0 | 0 |
| 18 | id: 6 dist: 25 | id:10 dist: 109 | id:15 dist: 155 | 2 | 10 | 0 | 10 | 0 | 10 | 0 |
| 2 | id:19 dist: 119 | id: 9 dist: 153 | id:12 dist: 163 | 1 | 19 | 0 | 19 | 0 | 19 | 0 |
| 3 | id: 8 dist: 72 | id:13 dist: 125 | id: 4 dist: 126 | 1 | 8 | 0 | 4 | 0 | 4 | 0 |
| 9 | id: 2 dist: 153 | id:14 dist: 187 | id: 1 dist: 216 | 1 | - | 0 | - | 0 | 19 | 0 |
| 12 | id: 4 dist: 112 | id:14 dist: 121 | id: 8 dist: 124 | 1 | 4 | 0 | 4 | 0 | 4 | 0 |
| 14 | id:12 dist: 121 | id: 2 dist: 186 | id: 9 dist: 187 | 1 | - | 0 | - | 0 | 4 | 0 |
| 15 | id:10 dist: 100 | id: 6 dist: 154 | id:18 dist: 155 | 1 | 10 | 0 | 10 | 0 | 10 | 0 |
| 16 | id: 0 dist: 115 | id: 1 dist: 167 | id:15 dist: 199 | 1 | 0 | 0 | 0 | 0 | 0 | 0 |
| 17 | id: 7 dist: 48 | id: 0 dist: 134 | id:18 dist: 177 | 1 | 0 | 0 | 0 | 0 | 0 | 0 |
| 11 | id: 4 dist: 131 | id:13 dist: 152 | id: 8 dist: 165 | 0 | 4 | 0 | 4 | 0 | 4 | 0 |

In the first step of the third phase (see Fig.1(c)) the peers 0, 4, 10, and 19 that have received a number of votes equal or greater than the $leader\_threshold$ value are elected actual leaders. It is worth to note that the peer 3 has not any actual leader. This because the node 8, which is a its potential leader, was not elected actual leader. Then the peer 3 locks among its neighbors to find an actual leader, and it selects the node 4. In case of more available neighbors the most similar one would be chosen. Moreover, the peer 9 and 14 did not find an actual leaders, since none of their neighbor peers is became actual leader. Such node are elaborated in the second step of this phase, where they look for an actual leader among the ones of their neighbor peers. As can be seen in the table, such nodes elect as their actual leader the node 19 and 4, respectively, which are neighbors of the peers 2 and 12. As final result (see Figure 1(e)) we obtain an overlay network made up of four communities composed by the peers 4 (actual leader), 3, 8, 11, 12 13 and 14 the first one, the peers 0 (actual leader), 7, 16, and 17, the second one, the peers 10 (actual leader), 6, 15, and 18, the third one, and the peers 19 (actual leader), 1, 2, 5 and 9, the though one.

## VI. Conclusions

In this paper we proposed AP2PLE, an asynchronous peer-to-peer leader election algorithm. Its aim is to identify, in a peer to peer fashion, a set of leaders representing user communities. A leader is a peer elected by a set of other peers as their representative on the basis of their similarity with it. The approach is based on peer-to-peer epidemic protocols for spreading information about network nodes. Each peer is associated with a network user, and is characterized by a profile. The leader's profiles is used as the community identifier. The election mechanism allows both the gathering into communities and the determination of the community leaders. Starting from the "private" vision of each node, after successive interactions and exchange of information among peers, it allows to exploit each peer local knowledge to contribute to the identification of the community leaders.

### Acknowledgement

We acknowledge the support of the CONTRAIL (EU-FP7-257438) and RECOGNITION (EU-FP7-257756) EU projects.